\documentclass[journal]{IEEEtran}

\usepackage{cite}
\usepackage[T1]{fontenc}
\usepackage{amsmath}
\usepackage{xspace}
\usepackage{amssymb} 
\usepackage{xcolor}
\usepackage{graphicx} 
\usepackage{subcaption} 
\usepackage{float} 
\usepackage{caption}
\usepackage{subcaption}

\newcommand{\D}{\bf{D}}

\ifCLASSINFOpdf

\else

\fi

\begin{document}

\title{GAN-driven Electromagnetic Imaging of 2-D Dielectric Scatterers}

\author{
        Ehtasham~Naseer,~\IEEEmembership{Graduate Student Member,~IEEE,}
        Ali Imran~Sandhu,~\IEEEmembership{Member,~IEEE,}
        Muhammad Adnan~Siddique,~\IEEEmembership{Senior~Member,~IEEE,}
        Waqas~W. Ahmed,~
        Mohamed~Farhat,~
        and~Ying~Wu
        
\thanks{Manuscript received XXX XX, 2024; revised XXX XX, 2024.}}

\markboth{February 2024}%
{Shell \MakeLowercase{\textit{NASEER et al.}}: GAN-DRIVEN EM IMAGING}

\maketitle


\begin{abstract} Inverse scattering problems are inherently challenging, given the fact they are ill-posed and nonlinear. This paper presents a powerful deep learning-based approach that relies on generative adversarial networks to accurately and efficiently reconstruct randomly-shaped two-dimensional dielectric objects from amplitudes of multi-frequency scattered electric fields. An adversarial autoencoder (AAE) is trained to learn to generate the scatterer's geometry from a lower-dimensional latent representation constrained to adhere to the Gaussian distribution. A cohesive inverse neural network (INN) framework is set up comprising a sequence of appropriately designed dense layers, the already-trained generator as well as a separately trained forward neural network. The images reconstructed at the output of the inverse network are validated through comparison with outputs from the forward neural network, addressing the non-uniqueness challenge inherent to electromagnetic (EM) imaging problems. The trained INN demonstrates an enhanced robustness, evidenced by a mean binary cross-entropy (BCE) loss of $0.13$ and a structure similarity index (SSI) of $0.90$. The study not only demonstrates a significant reduction in computational load, but also marks a substantial improvement over traditional objective-function-based methods. It contributes both to the fields of machine learning and EM imaging by offering a real-time quantitative imaging approach. The results obtained with the simulated data, for both training and testing, yield promising results and may open new avenues for radio-frequency inverse imaging.
\label{abstract}
\end{abstract}


%
\IEEEpeerreviewmaketitle

\section{Introduction}
\label{sec:Introduction}

\IEEEPARstart{T}{he} electromagnetic (EM) scattering problem is generally formulated using the volume integral equation (VIE) \cite{pastorino_book}, and its solution is typically determined via an iterative optimisation procedure \cite{sandhu_pasd,sandhu_bed}, in which the significant computational burden is often attributed to the repeated evaluations of the forward solver during each iteration. The inverse scattering problem is ill-posed, which renders the inversion sensitive to measurement noise and can cause instabilities. The inherent non-linearity further adds to the complexity of the problem. There are several persisting challenges; but as a whole, the field of EM imaging is presently undergoing a significant transformation, from the conventional numerical modeling approaches to the newer deep learning (DL) approaches that are proving to be an efficient way to solve the complex forward and inverse scattering problems \cite{D4,D5,D7,A28,A29}.

\subsection{Related Work}
DL-driven EM imaging solutions can be classified into two main streams: First, there are indirect schemes that (i) use images from first-order approximations like back propagation and feed them to a convolutional neural network (CNN) to get more accurate and refined images \cite{backpropag_cnn}; and those that (ii) leverage simple feed-forward neural networks to generate auxiliary parameters that are then fed to more complex algorithms to predict inverse parameters \cite{sandhu_cosamp}. However, this later technique has only been successful for investigation domains having low contrast relative to the background medium. A notable approach is replacing or augmenting the traditional numerical forward solver with a pre-trained neural network, acting as a surrogate model, offering potential reductions in the overall computational burden. Along these lines, recently, physics-embedded ML solutions \cite{D,A} have also demonstrated potential for overcoming the limitations of conventional computation-driven methods. In \cite{D}, a trained Born iterative method based on a U-net architecture is shown to accelerate convergence of the algorithm and low memory utilization throughout the training process.  In \cite{A}, the conjugate gradient method, which is often used to solve matrix systems, is turned into an iterative deep neural network to make the VIE solver work faster. In order to accelerate the computation of induced currents in the method of moments, the authors in \cite{GANpaper} have incorporated the `pix2pix' model \cite{GAN36}, which is a variant of Generative Adversarial Networks (GANs) \cite{GAN37}. Instead of directly solving the matrix equation, the proposed scheme learns the mapping between the unknown induced current and the given inputs. It then computes the scattered fields by multiplying these predicted currents with Green's function. However, both \cite{A,GANpaper} addressed the forward solution only.

The second class of methods are termed direct schemes that make use of DL to directly retrieve material properties from the measured scattered fields \cite{D1, D2, D3}. This is computationally more intensive and challenging due to the inherent ill-posedness associated with the EM inverse operator \cite{pastorino_book}. The one-to-many mapping limits the accuracy of the inverse design process. DL networks can learn nonlinear mappings in datasets, which enables them to approximate the true solution for on-demand inverse designs. In the field of acoustics, multiple studies have focused on discriminative and generative designs, which have demonstrated excellent results \cite{W18,W19,W20,ahmed2021deterministic,ahmed2023machine}. A linear sampling technique (LSM) integrated with a neural network to accurately reconstruct acoustic far-field data is introduced in \cite{W21}. Since LSM involves picking a contour to determine the support of the scatterer, the solution still offered some imprecision and requires further improvement. Furthermore, DL techniques based on probability density have also been explored to address the inverse problem in the context of acoustic meta-surfaces \cite{W24} and cloaks \cite{W25,W26}. Nevertheless, the accuracy of the proposed solutions is constrained to simple geometries, rendering the scaling of complex models challenging as the dimension of the design space increases. In this scenario, generative models are utilized to decrease the complexity of the design space and effectively discern the intricate connections between design parameters and system responses \cite{W19,W27}. In \cite{W29}, Wasserstein generative adversarial network is trained to reconstruct two-dimensional (2-D) rigid cylinders, given the total scattering cross sections. The scheme turned out to have a limited success in reducing the non-uniqueness, which drastically decrease the performance of the reconstruction algorithm in the far-field scattering problem. The recently proposed GAN-assisted inverse design process using multi-frequency acoustic phase-less far-field data \cite{W}, while promising, strategically simplified the problem. This simplification involved the use of a single plane wave source and the assumption that the far-field patterns would be smooth and invariant under the translation of the scattering object, thus avoiding the complexities of more intricate scenarios.

\subsection{Aim of this paper}
The aim of this paper is to leverage the capabilities of GANs for advancing EM imaging. More precisely, the focus lies in accurately and efficiently reconstructing 2-D dielectric scatterers incorporating multi-illumination and multi-frequency scattered electric fields. Once trained, the proposed DL-based framework provides a means to solve the inverse problem in real-time, thereby avoiding the large computational burden of the physics-based iterative numerical solvers.

\subsection{Outline}
In order to achieve the aforementioned aim, this study -- while building upon the earlier work in \cite{W} -- proposes a framework revolving around the training of three distinct DL models in a supervised fashion, each designed to address specific aspects of the problem at hand. Firstly, an adversarial autoencoder (AAE) is meticulously trained to generate 2-D dielectric scatterers from a latent space conditioned to be sampled from a Gaussian distribution. The training process spans $30,000$ epochs, employing Binary Cross-Entropy (BCE) loss function, ensuring effective optimization. Secondly, a CNN-based Forward Neural Network (FNN) is developed to establish the relationship between 2-D dielectric objects and their corresponding scattered electric fields at the four specified frequencies. To effectively optimize this CNN-based regression model while avoiding overfitting, we employed the mean squared error (MSE) as a loss function along with the L2 norm. Moreover, early stopping is implemented in response to a lack of improvement in the test loss, ensuring optimal model performance. Lastly, an inverse neural network (INN) is formulated, comprising three essential modules: an \emph{initially} untrained dense neural network, the already-trained generator pulled from the AAE, and the already-trained FNN. The untrained neural network processes the scattered electric fields, learning a latent space conforming to a Gaussian distribution. This latent space is then employed to generate 2-D dielectric scatterer profiles using the trained generator. To mitigate the non-uniqueness inherent in inverse design problems, the generated 2-D dielectric scatterer is fed into the trained FNN. In order to ensure that the latent space follow a Gaussian distribution and generated 2-D scatterer correspond to input scattered electric fields (ensuring unique solution), a combination of absolute loss and Kullback-Leibler (KL) loss functions is utilized to optimize the model.
\section{Problem Definition}
\label{sec:ProblemDefinition}
%
%
Let the investigation domain $\mathcal{D}$, characterized by the material properties $\left(\varepsilon(\mathbf{r}) = \varepsilon_{\text{r}}, \mu(\mathbf{r}) = \mu_0\right)$, be residing in a homogeneous background medium defined by $\left(\varepsilon(\mathbf{r}) = \varepsilon_0, \mu(\mathbf{r}) = \mu_0\right)$.
\begin{figure}
\centering
	{\includegraphics[width = 0.95\columnwidth]{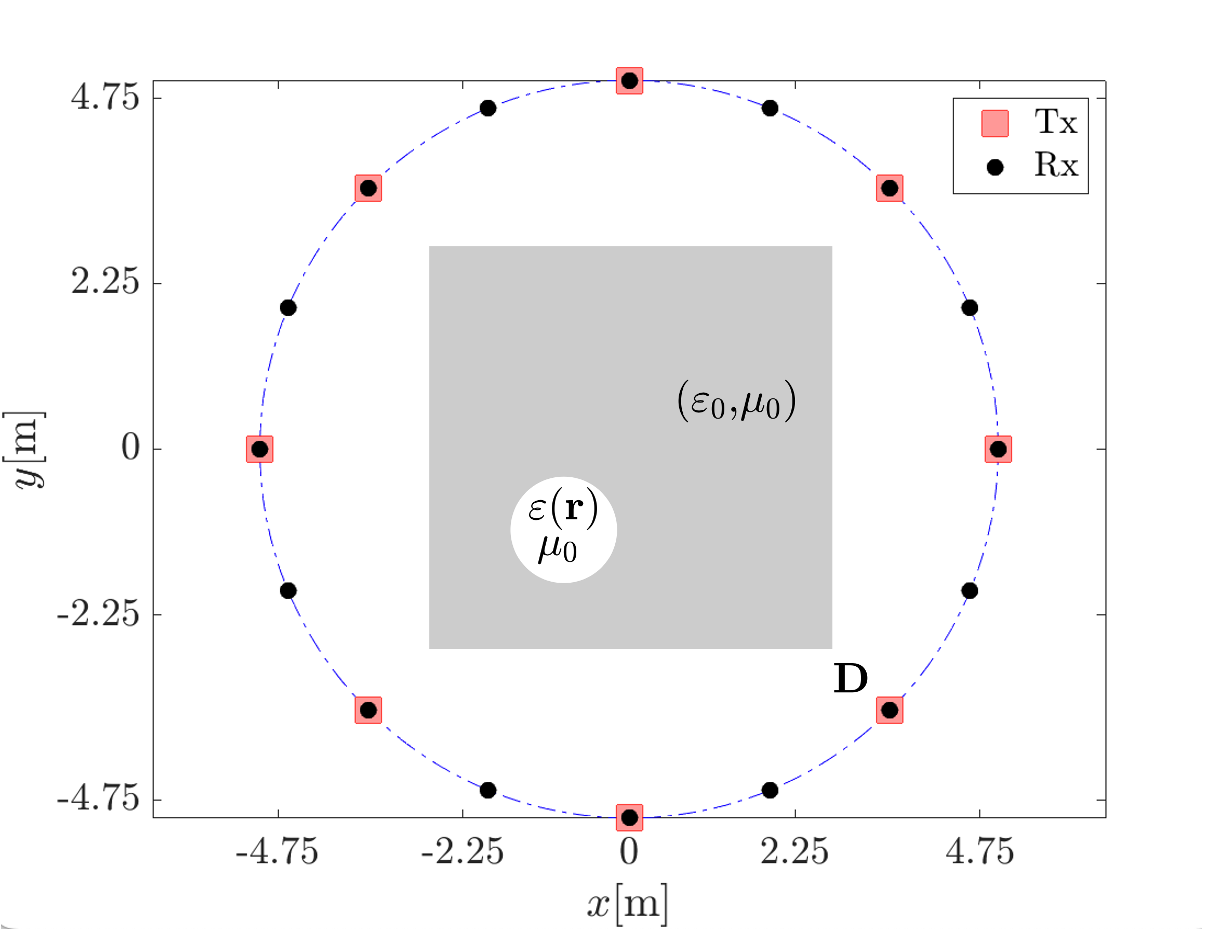}}
	\caption{Description of the 2-D EM imaging problem. Tx (Red) and Rx (Black) represents the transmitters, and receivers respectively. There are 8 Tx and 16 Rx used in the numerical simulations.}
	\label{fig1:EM}
\end{figure}
It is also  assumed that the material properties within $\mathcal{D}$, are invariant along the $z$-coordinate. There are $n_t$ transmitters located at $\mathbf{r}_i^{\mathrm{t}}, i=1, \ldots, n_t$ illuminating $\mathcal{D}$ with a $z-$polarized transverse-magnetic incident electric field $E_{i}^{\mathrm{inc}}({\mathbf{r}})$, one at a time, and $n_r$ receivers located at $\mathbf{r}_m^{\mathrm{r}}, m=1, \ldots, n_r$ sampling the scattered electric field $E_{i}^{\mathrm{sca}}({\mathbf{r}})$ as shown in Fig. \ref{fig1:EM}. Upon the illumination, secondary currents are induced within $\mathcal{D}$, resulting in the generation of the total electric field $E_{i}^{\mathrm{tot}}({\mathbf{r}})$ inside $\mathcal{D}$. The scattered electric field is related to the unknown contrast  $\tau{(\bf{r})} = \varepsilon({\mathbf{r}})/\varepsilon_0  - 1$ through the volume integral equation \cite{richmond1965}:
\begin{equation}
\label{eq_VIE}
E_{i}^{{\mathrm{sca}}}({\mathbf{r}}) = - {k_0^2}\int\limits_{\mathcal{D}} {\tau ({\mathbf{r'}})E_{i}^{{\mathrm{tot}}}({\mathbf{r'}})G({\mathbf{r}},{\mathbf{r'}})d\mathbf{r'}}.
\end{equation}   
Here, ${k_0} = \omega \sqrt {{\varepsilon _0}{\mu _0}} $  is the wave-number in the homogeneous background medium, $\omega = 2\pi f$ and $G({\mathbf{r}},{\mathbf{r'}}) = \frac{1}{4j} H_0^2(k_\mathrm{0} \left| {\mathbf{r}} - {\mathbf{r'}}\right|)$ is the 2-D scalar Green's function. Furthermore, $f$ is the frequency of the illuminating source, $j^2=-1$, and $H_0^2$ is the Hankel function of the second kind and order $0$. In (\ref{eq_VIE}) both $\tau{(\bf{r})}$ and $E_{i}^{\mathrm{tot}}({\mathbf{r}})$ are unknown, where $E_{i}^{\mathrm{tot}}({\mathbf{r}})$ constitutes the forward solver whose unique solution satisfies:
\begin{equation} 
\label{eq_Etot}
{E_{i}^{{\rm inc}} {(\bf{r})}=E_{i}^{\rm tot}{(\bf{r})}+{\it k}_{\rm 0}^{2} \int\limits _{\mathcal{D}}\tau {(\bf{r'})}E_{i}^{\rm tot} {(\bf{r'})}G{(\bf{r},\bf{r'})}d\mathbf{r'}}.
\end{equation}
%
To numerically solve equations \eqref{eq_VIE} and \eqref{eq_Etot},  $\mathcal{D}$ is discretized into $N$ square cells, each represented as $\mathcal{D}_n$, having centers ${\mathbf{r}_n}$, $n = 1, \ldots ,N$, upon which, the unknowns $\tau {(\bf{r})}$ and $E_{i}^{\rm tot}{(\bf{r})}$ are expanded using the pulse basis functions \cite{richmond1965,sandhu_bed}. The resulting equations are evaluated within and outside $\mathcal{D}$ resulting in the discretized system \cite{sandhu_bed}:
%

\begin{align}
\label{eq_sca_matrix}
\bar{E}_{i}^{{\text{sca}}} = -\bar{G}^{\text{R}} \hbox{diag}\left( \bar{E}_{i}^{\text{tot}}\right) \bar{\tau} = \bar{H}_i \bar{\tau}.
\end{align}
In \eqref{eq_sca_matrix}:
\[
\boldsymbol{\bar{E}}_i^{\text{tot}} = \begin{bmatrix} E_{i}^{\text{tot}}(\mathbf{r}_1) \\ E_{i}^{\text{tot}}(\mathbf{r}_2) \\ \vdots \\ E_{i}^{\text{tot}}(\mathbf{r}_N) \end{bmatrix},
%
%
\boldsymbol{\bar{E}}_i^{\text{sca}} = \begin{bmatrix} E_{i}^{\text{sca}}(\mathbf{r}_1^{r}) \\ E_{i}^{\text{sca}}(\mathbf{r}_2^{r}) \\ \vdots \\ E_{i}^{\text{sca}}(\mathbf{r}_{N}) \end{bmatrix}
\]
$\bar{\tau} = \left[ \tau(\mathbf{r}_1), \tau(\mathbf{r}_2), \cdots, \tau(\mathbf{r}_N)\right]^\top$,  and the entries of the matrices $\bar G^{\D}$ and ${\bar{G}^{\text{R}}}$ are: 
\begin{eqnarray*}
  \bar{G}^{\D}_{n,n'} = k_0^2 \int \limits_{D_{n'}} {G({{\mathbf{r}}_n}, {\mathbf{r'}})d\mathbf{r'}}, \quad
  \bar{G}^{\text{R}}_{m,n'} = k_0^2\int\limits_{D_{n'}} {G({\mathbf{r}}_m^{r},{\mathbf{r'}})d\mathbf{r'}}.
\end{eqnarray*} 
\begin{figure*}[!h]
  \centering
  \includegraphics[width = 0.95\textwidth]{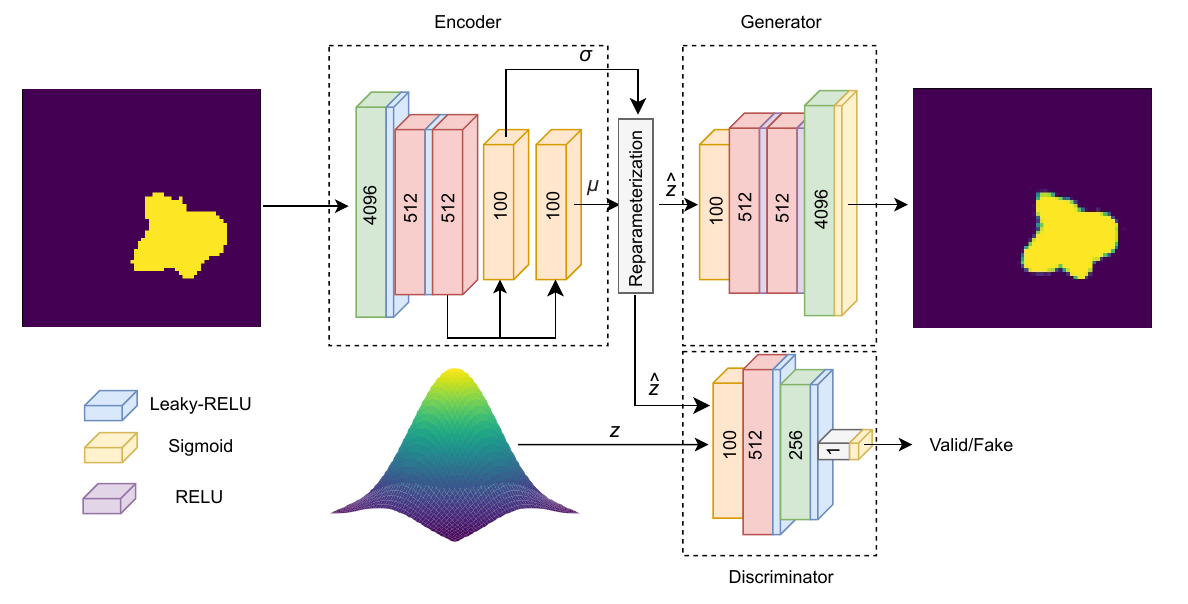}
  \caption{Architecture of the adversarial autoencoder (AAE) designed for generating the structure of physical object by learning a predefined distribution. In the encoder phase, both the variance ($\sigma$) and mean ($\mu$) of the specified distribution are generated, and these parameters are subsequently re-parameterized to produce the latent space ($\hat{z}$). The generator part then utilizes $\hat{z}$ to reconstruct the original image. Simultaneously, the discriminator is trained to distinguish between samples drawn from the predefined distribution, $z$, and those generated by the model, $\hat{z}$.}
  \label{fig:AAE}
\end{figure*}
While the primary focus of this paper is not on providing a numerical solution for the inverse EM problem, for the sake of completeness, it is noteworthy to mention that \eqref{eq_VIE}-\eqref{eq_sca_matrix} are derived assuming a single excitation. 
In cases involving multiple transmitters, these equations can be seamlessly integrated into larger matrix systems:
\begin{align}
\label{eq_sca_system}
\bar{E}^{{\text{mea}}} = \bar{H} \bar{\tau}.
\end{align}
Here, $\bar{E}^{\text {meas }} \approx \bar{E}^{\text {sca }}+\bar{\eta}$, where $\bar{\eta}$ represents the additive white Gaussian noise, and $\bar{E}^{\text {sca }}$ and $\bar{H}$ are obtained by respectively cascading $\bar{E}_i^{\text {sca }}$ and $\bar{H}_i$ for all $i=1, \ldots, n_{\mathrm{t}}$. The Matrix system (\ref{eq_sca_system}) exhibits nonlinearity with respect to $\bar{\tau}$ and is also ill-posed. In order to obtain an adequate solution, it is necessary to incorporate a regularisation scheme, as outlined in \cite{pastorino_book}. This involves solving the following optimization problem:
\begin{equation}
\label{eq_em_opt}
\bar{\tau}^*=\underset{\bar{\tau}}{\arg \min } \Bigl\{ \frac{1}{2}\|\bar{E}^{\text {meas }} -\bar{H} \bar{\tau}\|_2^2 +\gamma\|\bar{\tau}\|_l \Bigr\}.
\end{equation}
Here, $\gamma$ represents the penalty term's weight and $l$ represents its norm. Depending upon if $\tau$ exhibits a sharp or a smooth (piece-wise continuous) profile $l \in\{0,1,2\}$ \cite{sandhu_cosamp,sandhu_pasd,sandhu_bed}. 

The objective of this paper is to extensively investigate the capabilities of GANs as a viable alternative to conventional numerical schemes for obtaining efficient and accurate solutions to equations (\ref{eq_sca_matrix}) - (\ref{eq_em_opt}). 
\section{Generative Deep Learning Framework}
This study aims to address the inverse modeling for an EM imaging system, i.e. to design a framework to reconstruct the 2-D permittivity profile from scattered electric fields. We propose a innovative deep learning based design, such that it directly tackles the challenges caused by the non-linearity and ill-posedness of the inversion problem. The proposed design comprises three distinct DL models, integrated into a cohesive whole, resulting in a fully-trained inverse model that can be used for real-time quantitative imaging. The following subsection describes the individual models of the framework.


\subsection{Adversarial autoencoder (AAE)}
An autoencoder, by definition, is a neural network which attempts to learn a representation of the input data, by first reducing the data to a meaningful representation in a lower dimension, and thereafter attempting to recover or reconstruct the original from the same representation. On the other hand, a GAN consists of two neural networks, one is called the generator (which learns to generate realistic samples from random data) and the other is called the discriminator (which learns to differentiate between the real inputs and the fake samples generated by the generator). The AAE combines elements from both the autoencoder and the GANs. 

In this work, the proposed AAE architecture is as shown in Fig. \ref{fig:AAE}. The constituent encoder, generator, and discriminator consist of $4096$-$512$-$512$-$100$-$100$, $100$-$512$-$512$-$4096$, and $100$-$512$-$256$-$1$ layers, respectively. The training technique of AAE combines the latent distribution learning of variational autoencoders \cite{W37} with the adversarial learning of generative adversarial networks (GANs) \cite{W36}. This enhances the efficiency of AAE networks since adversarial learning prioritizes the representation within the latent space rather than the output generation. More precisely, the encoder takes an input geometry and transforms it into a dense (compressed), continuous design space called a \emph{latent space} compelled to adhere to a specified prior distribution through adversarial means by the discriminator. Given the latent space, the generator then generates several designs. Each of the three neural networks undergoes updates individually by optimizing its respective loss function.
%

The discriminator's weights are updated by minimizing the following loss function:
\begin{equation}
 \mathcal{L}_{\mathrm{disc.}}=-\log \left(\widehat{\gamma}_{\mathrm{prior}}\right)-\log \left(1-\widehat{\gamma}_{\mathrm{latent}}\right).   
\end{equation}
Here, $\widehat{\gamma}_{\mathrm{prior}}$ refers to the prediction made for a sample drawn from the prior distribution, which is supposed to be 1. On the other hand, $\widehat{\gamma}_{\mathrm{latent}}$ reflects the prediction made for a sample obtained from an encoded input in the latent space, which is expected to be 0. This technique allows the discriminator to differentiate between whether the sample comes from the encoder's latent space or a randomly generated distribution.
%

The weights of the autoencoder are updated through the minimization of a composite loss function: 
\begin{equation}
  \mathcal{L}_{\mathrm{auto-enc.}}=-\log \left(\widehat{\gamma}_{\mathrm{latent}}\right)-\mathcal{L}_{\text {recon}}.  
\end{equation}
The first term serves to compel the autoencoder to deceive the discriminator, generating sample-encoding vectors resembling those from a prior distribution. This influences the discriminator's prediction, aiming for $\widehat{\gamma}_{\mathrm{latent}}$ to approach 1.
The second term accounts for the reconstruction error, which evaluates how accurately the autoencoder reconstructs the arbitrary input object from the latent space. It is defined as:
\begin{equation}
 \mathcal{L}_{\text {recon. }} = \frac{1}{N_{\textrm{pixels}}} \sum_{k=1}^{N_{\textrm{pixels}}} \gamma_i \log \left(\widehat{\gamma}_i\right)+\left(1-\gamma_i\right) \log \left(1-\widehat{\gamma}_i\right).   
\end{equation}
Here, $N_{\textrm{pixels}}$ represents the total number of pixels in the input as well as the output, while both $\gamma_i$ and $\widehat{y}_i$ take values in $\{0,1\}$, classifying if a pixel sample happens to fall outside or within $\mathcal{D}$, respectively. For example, $\widehat{y}_i \rightarrow 1$ shows that the $i^{\textrm{th}}$ pixel has a higher likelihood that it belongs to $\mathcal{D}$.

Once trained, the generator is extracted from the AAE and integrated in the inverse design. Given a vector from the latent space, the trained generator can generate designs in real space. In principle, if a scattered field at the input of the stand-alone generator results in correct reconstruction of the 2-D permittivity profile, the inversion design is complete. However, the inversion may not be unique. 

The following subsection introduces the second DL model of the proposed framework, which is akin to a forward solver. This model will be integrated in the inversion design to address the non-uniqueness problem.

\subsection{Forward neural network (FNN)}
\label{subsec: FNN}
\begin{figure*}[t]
  \centering
  \includegraphics[width = 0.95\textwidth]{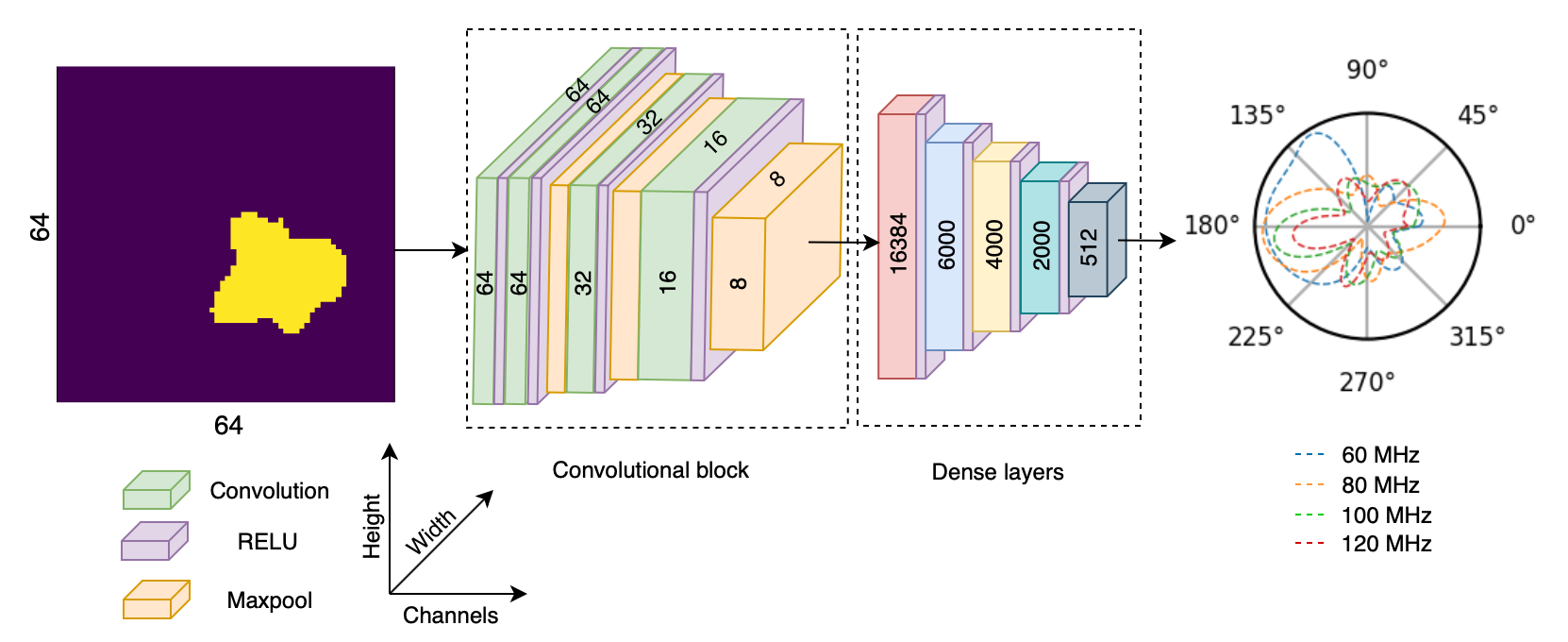}
  \caption{Architecture of the forward neural network (FNN), illustrating the spatial dimensions where the $z$-axis and $y$-axis represent the height and width of each layer, respectively. The $x$-axis corresponds to the number of channels in the corresponding layer.}
  \label{fig:fnn}
\end{figure*}
The primary purpose of integrating an FNN is to authenticate the images generated by the AAE. This is particularly critical because the inherent non-uniqueness in the generation process can lead to any image from the NULL space satisfying the input electromagnetic scattering response. Consequently, ensuring the accurate training of the FNN becomes vital for robust validation of these generated images.
\par 
Multiple deep neural networks were rigorously trained and tested to elucidate the intricate correlation between arbitrary 2-D dielectric profiles and their associated multi-frequency scattered electric fields, synthetically generated through the numerical solution of Eq. (\ref{eq_sca_matrix}). The training methodology adhered to a supervised approach, wherein the FNN was fed with $64\times64$ images of dielectric profiles as inputs, while the corresponding absolute values of complex-valued multi-frequency scattered electric fields served as labels. The outcomes of the experimentation revealed that simpler models tended to exhibit signs of underfitting, whereas more complex models demonstrated a tendency for overfitting. A thorough insight into the data unveiled inherent challenges, as the labels, representing the scattered electric fields, exhibited abrupt transitions and irregularities, making it challenging for a deep neural network to learn a discernible trend or structure. Subsequently, a CNN-based FNN was developed, leading to a noteworthy enhancement in accuracy. The architecture of the FNN, depicted in Fig. \ref{fig:fnn}, is composed of two principal blocks: a convolutional block and dense layers. The convolutional block comprises five layers, each utilizing a $3\times3$ filter with a $1\times1$ strid. It is followed by the Rectified Linear Unit (ReLU) activation function. Maxpool layers, featuring a $2\times2$ size with a $2\times2$ stride, are integrated to downsample the features extracted by the convolutional layers. Importantly, the number of channels expands by a factor of 2 after each convolution layer, augmenting the extraction of representative information from the salient features.
\par
 In Fig. \ref{fig:fnn}, the dimensions of each layer are denoted as $H\times W\times C$, which represent the height, width, and number of channels in a layer, respectively. The final layer of the convolutional block attains dimensions of $8\times8\times256$. This layer is flattened to serve as input for the second block, which encompasses dense layers with sizes $16,384$, $6,000$, $4,000$, $2,000$, and $512$. Each layer in this block employs a ReLU activation function, except for the ultimate output layer, which produces the predicted scattered electric fields, forming a vector of size 512.
\par
 To quantitatively evaluate the model's performance and fine-tune it through backpropagation, we incorporated mean squared error (MSE) with L2 regularization to mitigate the risk of overfitting. MSE calculates the average of squared differences between the actual and predicted scattered electric field, as follows:

\begin{equation}
\label{eqn_mse}
\mathcal{L}_{MSE} = \frac{1}{n} \sum_{i=1}^{n} (y_i - \hat{y}_i)^{2}.
\end{equation}
Here, $n$ represents the no of samples of the scattered electric field, $y$ denotes the true labels, and $\hat{y}$ signifies the predicted scattered electric field samples.
\subsection{Inverse neural network (INN)}
\begin{figure*}[t]
  \centering
  \includegraphics[width = 1.0\textwidth]{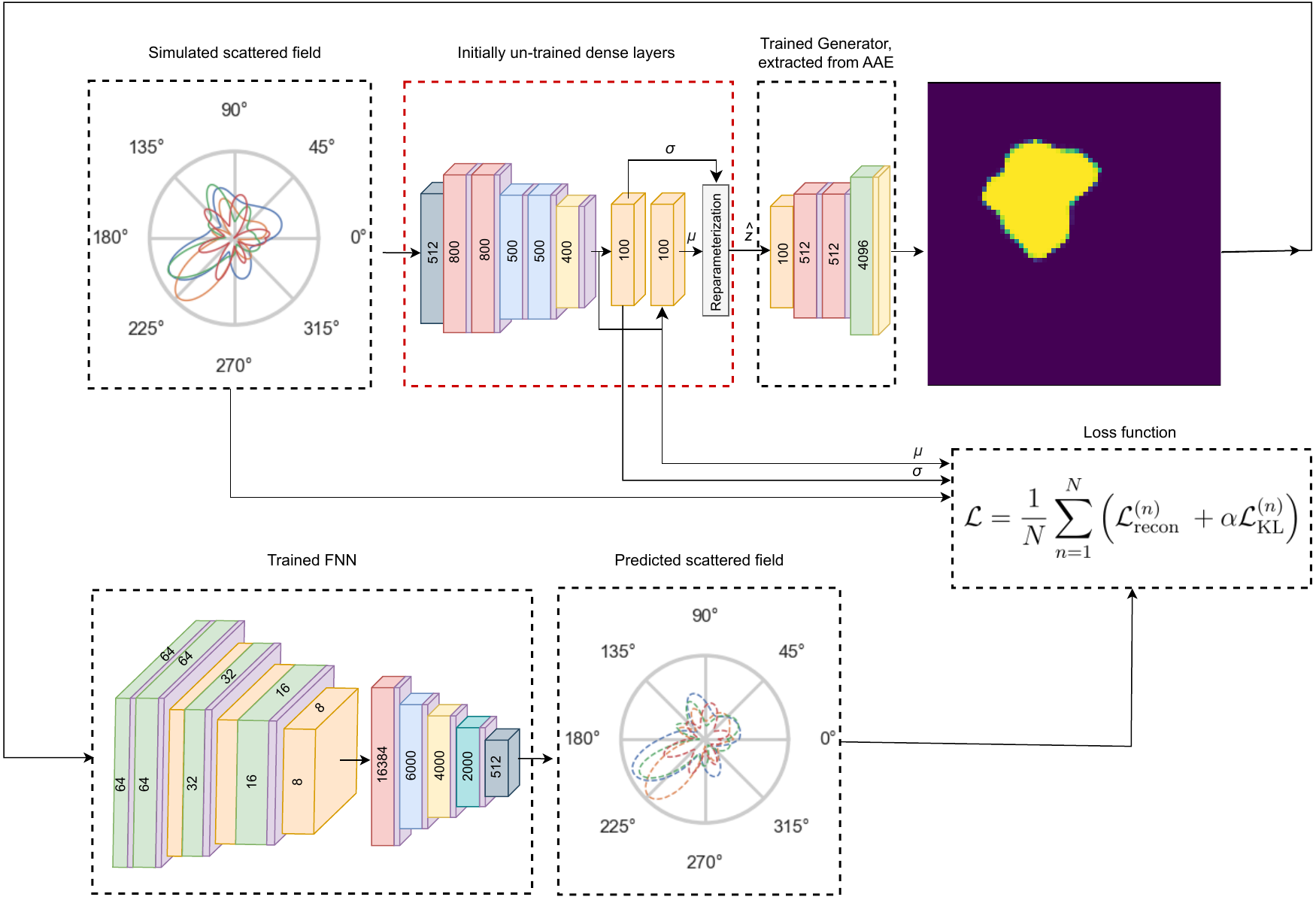
  }
  \caption{Schematic representation of complete architecture employed in developing the inverse neural network (INN). The INN integrates a dense layer block with a trained generator, forming a cohesive framework for inverse design. Trained FNN is only employed to facilitate the unique solution corresponding to scattered field. A combination of reconstruction loss and KL divergence is utilized for effective optimization of the model.}
  \label{fig:inn}
\end{figure*}
 Unlike forward modeling, which establishes an unambiguous mapping between a 2-D dielectric object and its corresponding scattered electric field, the inverse design struggles with the challenges arising from many-to-one mapping. Consequently, relying on a single deep learning model that provides the mapping from the response space to the structural space is deemed insufficient. In order to tackle the non-uniqueness of inverse design problem, forward solver is incorporated in the inverse design problem, leading to a tandem network. Conventional tandem networks struggle with acquiring the pixel-binary structure because of the considerable laxity in parameter design, making this learning process nontrivial. Moreover, The inherent complexity of the mapping task, coupled with the intricate nature of parameter design, poses difficulties in achieving a straightforward and complete elimination of non-uniqueness within the conventional tandem network framework. 
 \par
 In this study, we propose an inverse design strategy that leverages the dimensionality reduction capabilities of generative models and the effectiveness of incorporating forward solvers to overcome non-uniqueness in inverse design problems. Fig. \ref{fig:inn} illustrates the comprehensive schematic of the proposed tandem network, comprising initially untrained dense layers, a trained generator extracted from AAE, and a trained FNN. The dense network consists of eight layers with sizes of $512, 800, 800, 500, 500, 400, 100,$ and $100$ respectively, utilizing the ReLU function, to capture the non-linearities. This block maps the scattered electric field to latent space $z$, conditioned to belong to a Gaussian distribution parameterized by mean $\mu$ and standard deviation $\sigma$. In Fig. \ref{fig:inn}, the dense layer network in conjunction with the trained generator, constitutes the INN. The integration of FNN with INN is for ensuring a unique solution in the inverse design. 

 For effective learning and optimization of the model, we define a loss function that combines the absolute loss between simulated and generated scattered electric fields, denoted as $\mathcal{L}_{\text {recon}}$, and the Kullback-Leibler (KL), $\mathcal{L}_{\mathrm{KL}}$, divergence between the latent space distribution and prior Gaussian distribution. The overall loss function for training INN is expressed as:
\begin{equation}
\mathcal{L}=\frac{1}{N} \sum_{n=1}^N\left(\mathcal{L}_{\text {recon }}^{(n)}+\alpha \mathcal{L}_{\mathrm{KL}}^{(n)}\right),
\end{equation}
where $n$ is a data sample among a total of N samples, and $\alpha$ is a weighting parameter balancing deterministic and generative learning. The reconstruction loss, $\mathcal{L}_{\text {recon }}^{(n)}$, is computed as:
\begin{equation}
\mathcal{L}_{\text {recon }}^{(n)}=\sum_{m=1}^M\left|y_m^{(n)}-\widehat{y}_m^{(n)}\right|,
\end{equation}
where $M$ is the length of the input scattered field data, and $y_i$ and $\widehat{y}_i$ are the simulated and predicted scattered fields, respectively. The KL divergence is expressed as follows:
\begin{equation}
\mathcal{L}_{\mathrm{KL}}^{(n)}=-\frac{1}{2} \sum_j^J\left(1+\left(\sigma_j^{(n)}\right)^2-\left(\mu_j^{(n)}\right)^2-\log \left(\sigma_j^{(n)}\right)^2\right),
\end{equation}
where $J$ is the length of the latent vector.

\begin{figure*}[!t]
    \centering
    \begin{subfigure}[b]{0.32\textwidth}
        \centering
        \includegraphics[width=\textwidth]{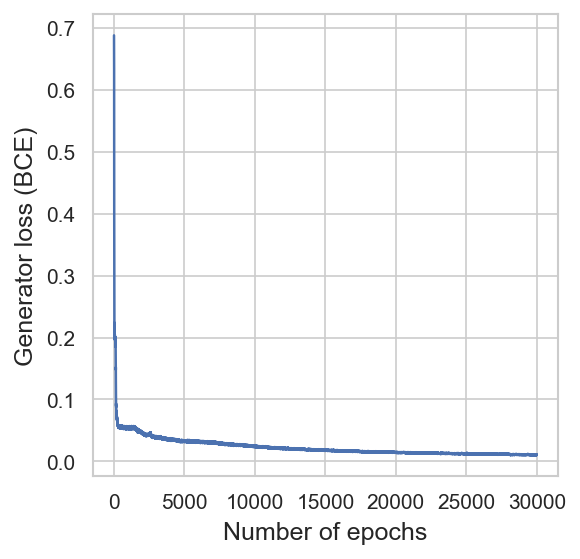}
        \caption{}
        \label{fig:gen_loss}
    \end{subfigure}
    \hfill
    \begin{subfigure}[b]{0.32\textwidth}
        \centering
        \includegraphics[width=\textwidth]{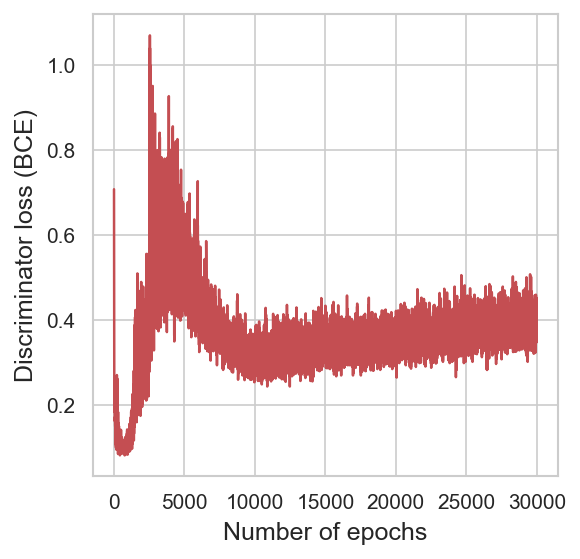}
        \caption{}
        \label{fig:dic_loss}
    \end{subfigure}
    \hfill
    \begin{subfigure}[b]{0.32\textwidth}
        \centering
        \includegraphics[width=\textwidth]{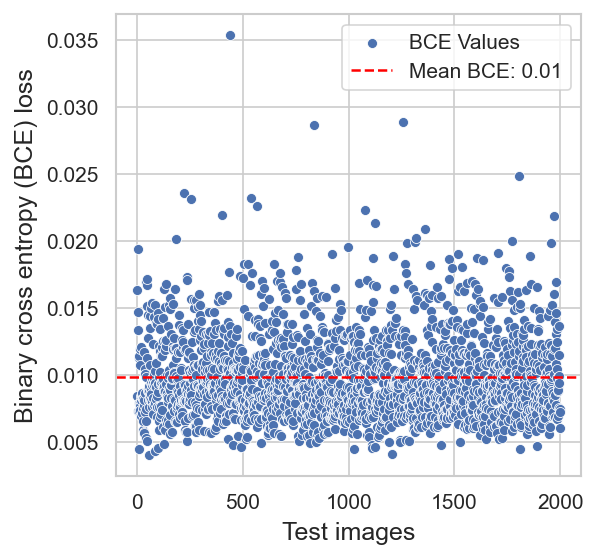}
        \caption{}
        \label{fig:AAE_BCE_loss}
    \end{subfigure}
    \newline
    \begin{subfigure}[b]{0.32\textwidth}
        \centering
        \includegraphics[width=\textwidth]{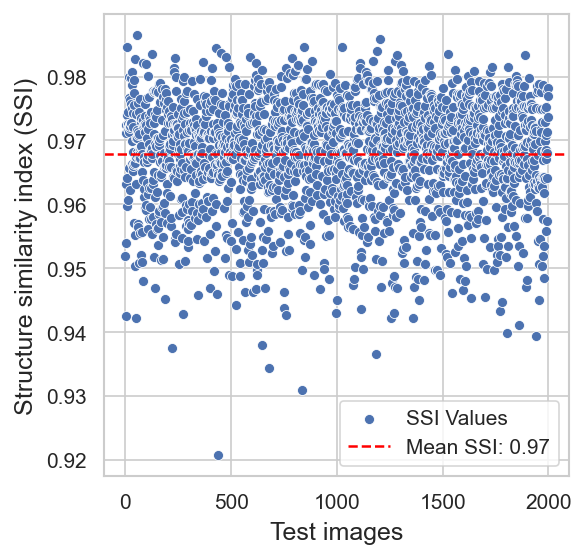}
        \caption{}
        \label{fig:AAE_SSIM}
    \end{subfigure}
    \hspace{1cm}
    \begin{minipage}[b]{0.32\textwidth}
        \centering
        \begin{subfigure}[b]{0.46\textwidth}
            \centering
            \includegraphics[width=\textwidth]{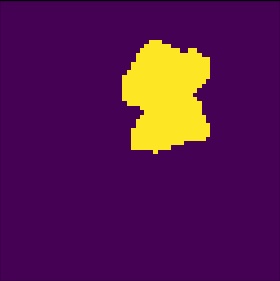}
            \caption{Original}
            \label{fig:sub5}
        \end{subfigure}
        \begin{subfigure}[b]{0.46\textwidth}
            \centering
            \includegraphics[width=\textwidth]{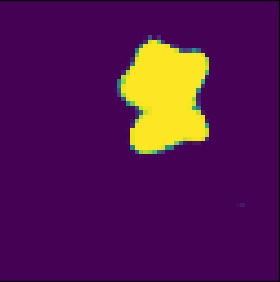}
            \caption{Predicted}
            \label{fig:AAE_generated_img1}
        \end{subfigure}
        \newline
        \begin{subfigure}[b]{0.46\textwidth}
            \centering
            \includegraphics[width=\textwidth]{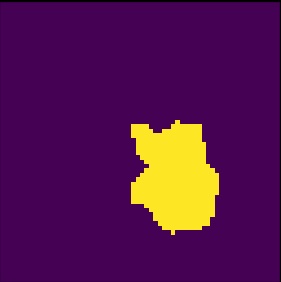}
            \caption{Original}
            \label{fig:sub7}
        \end{subfigure}
        \begin{subfigure}[b]{0.46\textwidth}
            \centering
            \includegraphics[width=\textwidth]{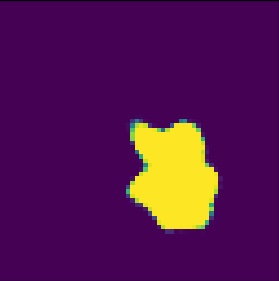}
            \caption{Predicted}
            \label{fig:AAE_generated_img2}
        \end{subfigure}
    \end{minipage}
\caption{Training and testing results of adversarial autoencoder (AAE). (a) Binary Cross Entropy (BCE) loss curve, illustrating the gradual and smooth convergence of the loss over successive epochs. (b) BCE loss curve representing the learning progress of discriminator over epochs. The fluctuations in the graph are attributed to adversarial learning, and ideally, it should remain close to $0.5$. (c) \& (d) Display scatter plots illustrating BCE loss and Structure Similarity Index (SSI) computed across the entire test dataset, comprising $2000$ images. The mean BCE value is observed to converge around $0.01$, while mean SSI remains around $0.97$. (e) \& (h) showcase the comparison of two randomly selected test images and corresponding generated images by AAE.}
\label{fig:AAE_results}
\end{figure*}
\section{Numerical Results \& Discusion}\label{sec:results}
This section presents the details related to the simulated data and the training of all the above mentioned deep learning models models along with the corresponding results. 
\subsection{Simulated data}
The 2-D permittivity distributions are modeled by generating synthetic images, corresponding to 9.45 x 9.45 meters, or $5
\lambda \times 5\lambda$ in dimensions.
$30,000$ dielectric scatterers have been simulated, while computing the scattered electric fields at four distinct frequencies ($60$ MHz, $80$ MHz, $100$ MHz, and $120$ MHz) utilizing an iterative (forward) solver \cite{sandhu_pasd,sandhu_bed}. The simulation set up follows the schematic shown in Fig. \ref{fig1:EM}, employing $8$ transmitters and $16$ receivers. This schematic is consistently applied across all four frequencies, resulting in a vector of scattered electric fields of size $128$ for each frequency. Subsequently, these four vectors of scattered electric fields are concatenated to form a unified vector of size $512$ for each 2-D dielectric scatterer. Arguably, given the fact that the frequency used in an EM imaging system varies from application to application, the choice of the aforementioned frequencies is non-specific, and hence it does not impede generality of the problem. 
\subsection{Training and testing results of AAE}
For the training of AAE, the dataset is partitioned into subsets of $26,000$, $2,000$, and $2,000$ samples for training, validation, and testing, respectively. The $64 \times 64$ input 2-D dielectric scatterer is flattened, making a vector of size $4096$, to input to the AAE. Employing a batch size of $100$ and the Adam optimizer with a learning rate of $0.0002$, the model underwent rigorous training.  The training dynamics are visualized in Fig. \ref{fig:gen_loss} and \ref{fig:dic_loss}, showcasing the convergence patterns of the generator and discriminator losses. Notably, the generator loss exhibits a smooth convergence over epochs, whereas the discriminator loss experiences fluctuations attributed to adversarial learning. Despite these fluctuations, the mean value hovers around $0.4$. Ideally, this value should be $0.5$, as the discriminator's task is to discern between real samples from a predefined distribution and the generated ones. To comprehensively assess the performance of trained AAE, we delve into the evaluation metrics of BCE loss and Structure Similarity Index (SSI). Fig. \ref{fig:AAE_BCE_loss} and \ref{fig:AAE_SSIM} present the results of these two evaluation metrics on test data. The BCE loss attains a mean value of $0.01$, affirming effective training. Simultaneously, the SSI demonstrates a high mean value of $0.97$, indicating the fidelity and robustness of our AAE in preserving structural features. These metrics collectively validate the proficiency and reliability of our AAE model.
Fig. \ref{fig:AAE_generated_img1} and \ref{fig:AAE_generated_img2} provide visual insights into the model's generative capabilities, depicting two example images generated by the trained AAE. Observing the generated objects and comparing with their respective test counterparts, it is evident that the generator effectively produces outputs that closely resemble the original test images.
%
%

\subsection{Training and testing results of FNN}
\begin{figure}
	\centering
	\begin{subfigure}{0.85\linewidth}
		\includegraphics[width=\linewidth]{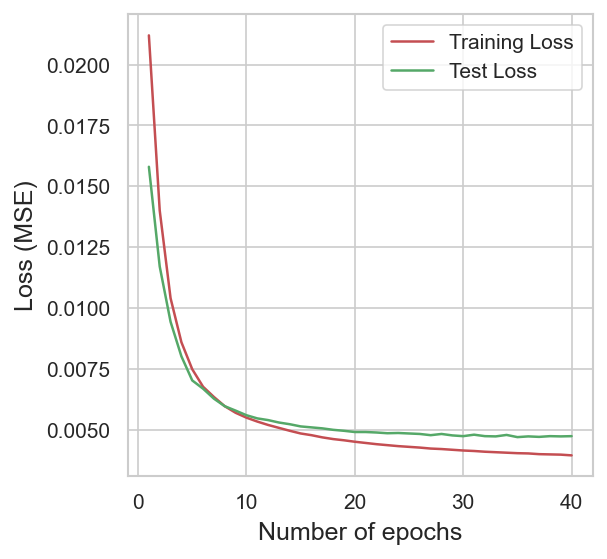}
		\caption{}
		\label{fig:fnn_curves_subfig}
	\end{subfigure}\\
	\begin{subfigure}{0.85\linewidth}
		\includegraphics[width=\linewidth]{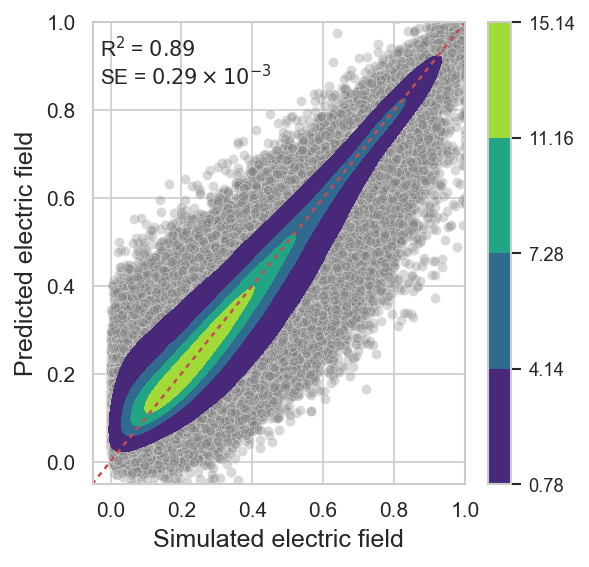}
		\caption{}
		\label{fig:fnn_regr_subfig}
	\end{subfigure}
	\caption{Training and test results of forward neural network (FNN). (a) Mean Squared Error (MSE) training and test loss curves. The optimal test MSE loss was observed around $0.46 \times 10^{-2}$. (b) Scatter plot illustrating the relationship between predicted and actual test values, accompanied by a linear regression fit shown in red dashed line. The yellowish-to-blue color palette represents 2-D kernel density highlighting regions where the points are more densely situated.}
	\label{fig:tranied_fnn_results}
\end{figure}

The main objective of FNN is to learn the underlying complex relation between 2-D dielectric scatterer and corresponding scattered electric field. For this objective, the 2-D dielectric of size $64 \times 64$ with $1$ channel is fed to the convolutional block of the FNN, with $27,000$ data samples as training and remaining $3,000$ as testing. During the training process, a batch size of $30$ and Adam optimizer with learning rate of $0.001$ is employed to ensure effective learning. Additionally, early stopping subjected to test loss is applied with a patience of $5$ to prevent overfitting and increase the generalizability of the model.  Fig. \ref{fig:fnn_curves_subfig} elucidates the gradual convergence of training and test loss, culminating in an early stoppage after $40$ epochs due to a lack of improvement in test loss. Notably, the model attains optimal test performance, achieving an average MSE of $0.46 \times 10^{-2}$. Moreover, the coefficient of determination (R\textsuperscript{2}) is computed to assess the model's efficacy in learning and explaining the variability within the data, yielding a value of $0.89$. Fig. \ref{fig:fnn_regr_subfig} showcases the relationship, through scatter plot, between simulated and predicted electric fields of all the test points along with a linear fitted regression line (depicted as red-dashed). The 2-D kernel density, presented in yellowish-to-blue color gradient, signifies varying point densities near the linear fit line. High point density near the linear fit line, a low value of standard error (SE) of $0.29 \times 10^{-2}$, and high R\textsuperscript{2} value of $0.89$ collectively signifies the similarity between the simulated and predicted values. 
\par
\begin{figure*}
	\centering
	\begin{subfigure}{0.3\textwidth}
		\includegraphics[width=\linewidth]{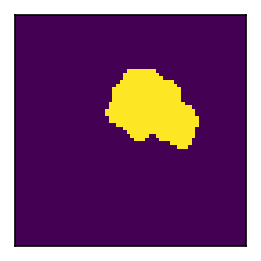}
		\caption{}
		\label{fig:fnn_input_img_subfig}
	\end{subfigure}
	\begin{subfigure}{0.32\textwidth}
		\includegraphics[width=\linewidth]{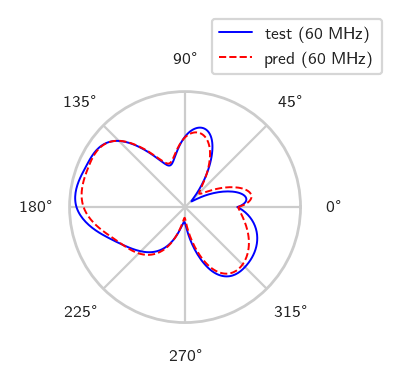}
		\caption{}
		\label{fig:fnn_polar60_subfig}
	\end{subfigure}
 	\begin{subfigure}{0.32\textwidth}
		\includegraphics[width=\linewidth]{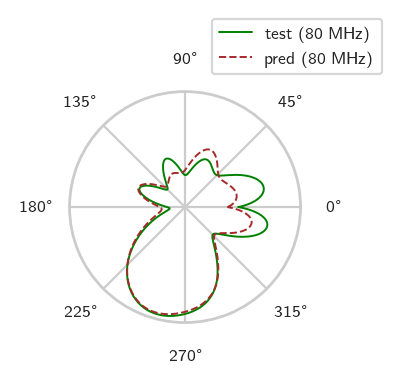}
		\caption{}
		\label{fig:fnn_polar80_subfig}
	\end{subfigure}\\
	\begin{subfigure}{0.32\textwidth}
		\includegraphics[width=\linewidth]{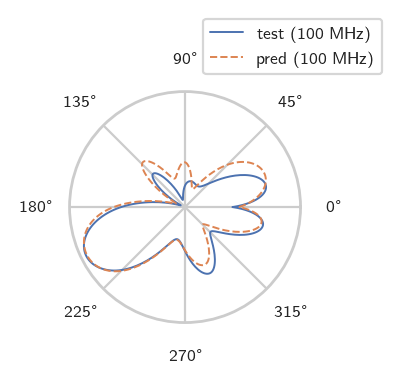}
		\caption{}
		\label{fig:fnn_polar100_subfig}
	\end{subfigure}
 	\begin{subfigure}{0.32\textwidth}
		\includegraphics[width=\linewidth]{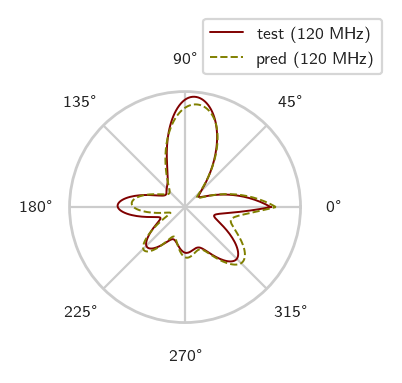}
		\caption{}
		\label{fig:fnn_polar120_subfig}
	\end{subfigure}
 
	\caption{Results from the trained forward neural network (FNN). (a) A randomly selected input image from the test dataset. Subfigures (b), (c), (d), and (e) present a comparative analysis of the test and predicted scattered electric field at frequencies of $60$ MHz, $80$ MHz, $100$ MHz, and $120$ MHz, respectively.}
	\label{fig:traind_fnn_results}
\end{figure*}

In Fig. \ref{fig:traind_fnn_results}, results for a randomly selected test point are presented. Fig. \ref{fig:fnn_input_img_subfig} illustrates the test input image, where the yellow region represents a randomly shaped 2-D dielectric object. Sub-figs. \ref{fig:fnn_polar60_subfig}, \ref{fig:fnn_polar80_subfig}, \ref{fig:fnn_polar100_subfig}, and \ref{fig:fnn_polar120_subfig} provide a comparative analysis between the simulated and predicted scattered electric field from a randomly selected transmitter's position. Upon examination of the graphs, it is apparent that the predicted values generally exhibit a pattern consistent with the variability observed in the simulated data. However, it is noteworthy that, in certain instances, the model predictions deviate from the true values, attributed to inherent errors in the model predictions. 

\subsection{Training and testing results of INN}
\begin{figure*}
	\centering
	\begin{subfigure}{0.33\textwidth}
		\includegraphics[width=\linewidth]{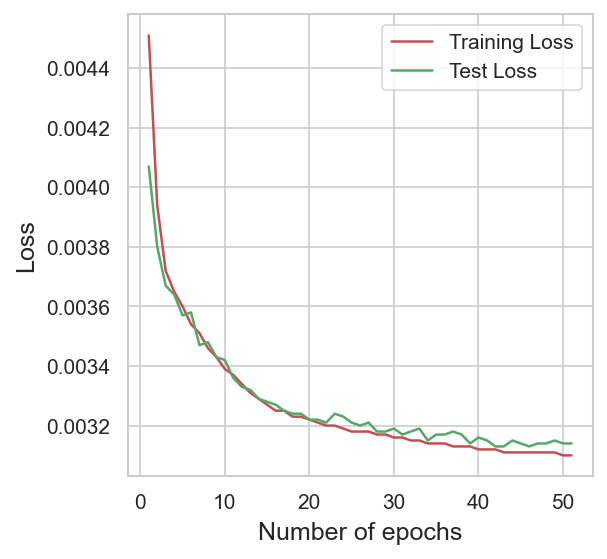}
		\caption{}
		\label{fig:inn_curves_subfig}
	\end{subfigure}
	\begin{subfigure}{0.32\textwidth}
		\includegraphics[width=\linewidth]{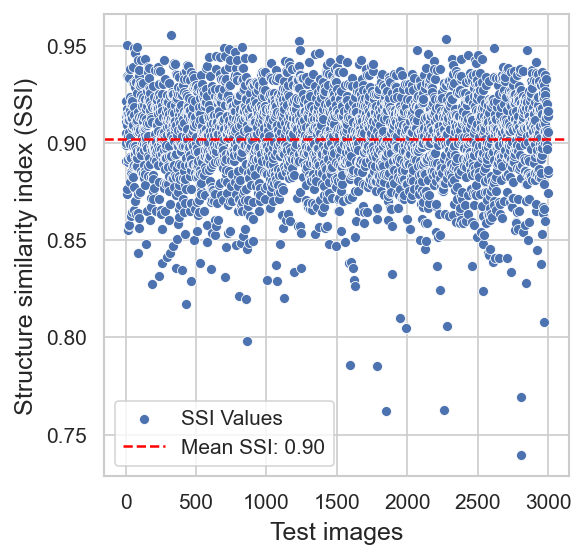}
		\caption{}
		\label{fig:inn_ssim_subfig}
	\end{subfigure}
 	\begin{subfigure}{0.32\textwidth}
		\includegraphics[width=\linewidth]{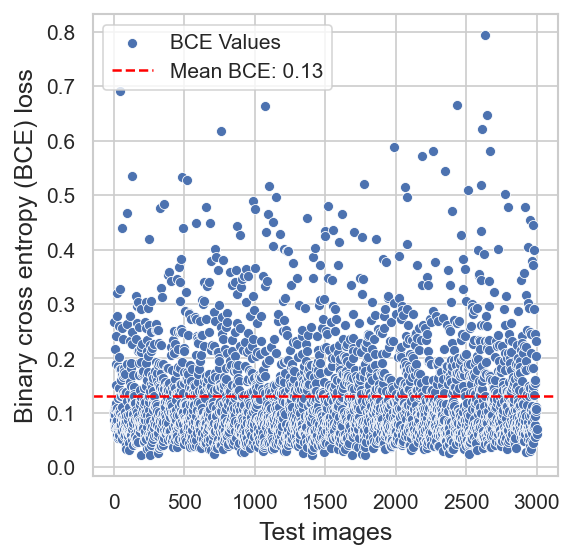}
		\caption{}
		\label{fig:inn_bce_subfig}
	\end{subfigure}\\
 
	\caption{Results of training and test metrics for the Inverse Neural Network (INN). Subfigure (a) illustrates the convergence of training and test loss curves over multiple epochs, with the best test loss observed at approximately $0.31 \times 10^{-2}$. Subfigure (b) presents a scatter plot depicting the Structure Similarity Index (SSI) for all $3000$ test points, accompanied by the mean value indicated by a red-dashed horizontal line. In Subfigure (c), a scatter plot showcases the Binary Cross Entropy (BCE) loss for all test points, with the mean value marked by a red-dashed horizontal line, observed to be approximately $0.13$.}
\label{fig:inn_model_results}
\end{figure*}

Fig. \ref{fig:inn} represents the complete schematic for the training of INN where the generator and FNN are already trained and only dense network is subjected to training. For the training of this network the dataset is partitioned into training ($27,000$ samples) and testing ($3,000$ samples) sets. The optimization process involves the Adam optimizer with a learning rate of $0.0002$, yielding satisfactory results. In fig. \ref{fig:inn_curves_subfig}, a gradual decrease in both training and test loss is observed, reaching an optimal test loss value of $0.31 \times 10^{-2}$ before triggering early stopping. Subsequently, the trained INN model, in conjunction with the pre-trained generator, is evaluated across the entire test dataset. Figs. \ref{fig:inn_ssim_subfig} and \ref{fig:inn_bce_subfig} present scatter plots depicting the SSI and BCE loss, respectively. The notably high mean SSI value of $0.90$ and low BCE loss value of $0.13$ affirm that the predicted and test images closely align both structurally and pixel-wise.
\par
\begin{figure*}
	\centering
	\begin{subfigure}{0.525\textwidth}
		\includegraphics[width=\linewidth]{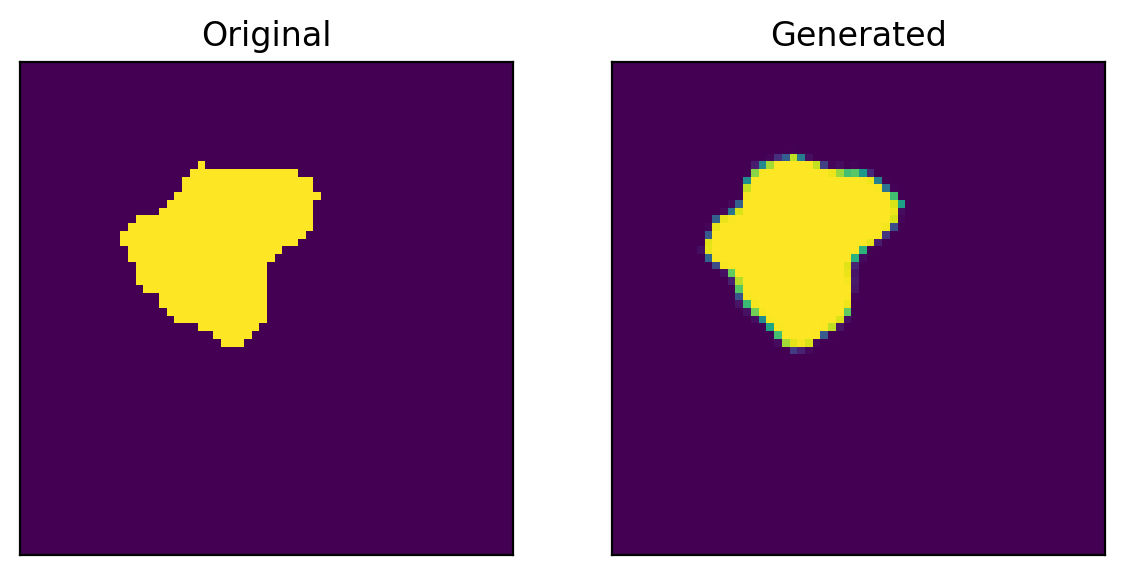}
        \label{fig:inn_test_img1_subfig}
	\caption{}
	\end{subfigure}
	\begin{subfigure}{0.32\textwidth}
	\includegraphics[width=\linewidth]{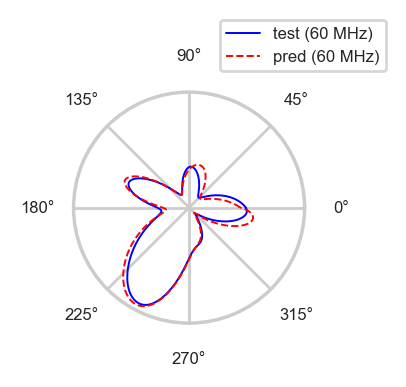}
 	\caption{}
        \label{fig:inn_polar60_subfig}
	\end{subfigure}\\
 	\begin{subfigure}{0.32\textwidth}
		\includegraphics[width=\linewidth]{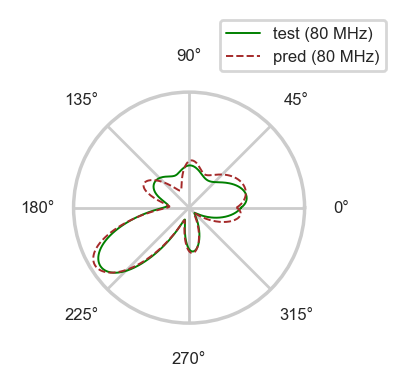}
  	\caption{}
        \label{fig:inn_polar80_subfig}
	\end{subfigure}
	\begin{subfigure}{0.32\textwidth}
		\includegraphics[width=\linewidth]{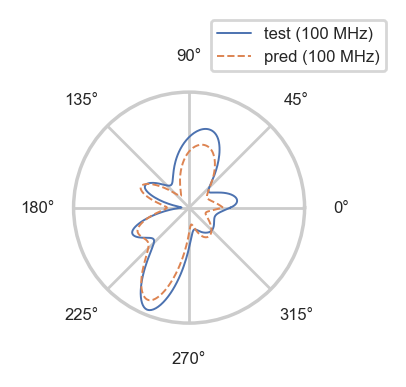}
  	\caption{}
        \label{fig:inn_polar100_subfig}
	\end{subfigure}
 	\begin{subfigure}{0.32\textwidth}
		\includegraphics[width=\linewidth]{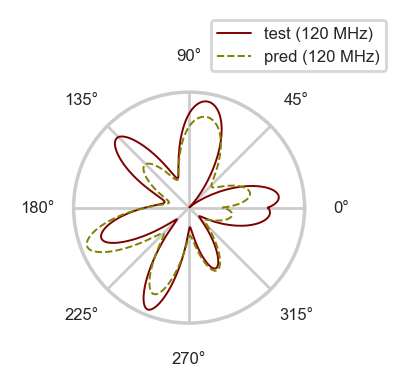}
  	\caption{}
        \label{fig:inn_polar120_subfig}
	\end{subfigure}
 
	\caption{Results from trained Inverse Neural Network (INN). The 'Original Image' represents a randomly selected test point from the test dataset, while the 'Generated Image' corresponds to the image produced by the INN when supplied with the scattered electric field of the same test point as input. Subfigures (a), (b), (c), and (d) depict the randomly selected back-scatter electric field for four different frequencies of a specific transmitter. These subfigures offer a comparative analysis between the input back-scatter electric field and the reproduced electric field by the model presented in Fig. \ref{fig:inn} for the generated object.}
	\label{fig:inn_example_results}
\end{figure*}

To visually assess the outcomes of the trained network, a representative test example is showcased in Figure \ref{fig:inn_example_results}. Notably, the generated image closely aligns with the corresponding test image, with only minor distortions observed at the edges of the generated image—attributed to inherent model errors. Furthermore, Figs. \ref{fig:inn_polar60_subfig}, \ref{fig:inn_polar80_subfig}, \ref{fig:inn_polar100_subfig}, and \ref{fig:inn_polar120_subfig} present a comparative analysis between the predicted and test scatter electric fields at frequencies of $60$ MHz, $80$ MHz, $100$ MHz, and $120$ MHz, respectively, for the same test image.These results demonstrate a consistent correspondence between the generated image and the input scattered electric field. 

\section{Conclusion}
\label{sec:Conclusion}
Solving the EM inverse scattering problem accurately and efficiently is a challenging task because (i) the scattered electric fields are a nonlinear function of the unknown dielectric profile, (ii) the measurements are finite and contaminated with noise, making the problem ill-posed, and (iii) the solution is typically determined via an iterative optimisation procedure, in which the significant computational burden is often attributed to the repeated evaluation.
\par
To address these challenges, the field of EM imaging is presently undergoing a significant transformation, driven by recent advances in deep learning approaches. However, the accuracy of the solutions is constrained to simple geometries, rendering the scaling of complex models challenging as the dimension of the design space increases. In this context, generative models are employed to reduce the dimensionality of the design space and efficiently learn the intricate relations between design parameters and system responses. 
\par
We proposed a generative deep learning approach as a practical design tool to uniquely determine the shape of an arbitrary object using multi-frequency scattered field data. To accelerate inverse learning, an adversarial autoencoder (AAE) is trained to learn to generate scatterer’s geometry from a lower-dimensional latent representation constrained to adhere to the Gaussian distribution. A cohesive inverse neural network (INN) framework is set up comprising a sequence of appropriately designed dense layers, the already-trained generator, and a separately trained forward neural network. The images reconstructed at the output of the inverse network are validated through comparison with outputs from the forward neural network, addressing the non-uniqueness challenge inherent in imaging problems. The designed network instantly predicts the shape of the object for the given scattered electric field information. It is important to emphasize that all our simulations are scalable, the frequency used in an EM imaging system varies from
application to application, the choice of the aforementioned
frequencies is non-specific, and hence it does not impede
generality of the problem.
\par
The study not only demonstrates a significant reduction in computational load but also marks a substantial improvement over traditional objective-function-based methods. It contributes both to the fields of machine learning and EM imaging by offering a real-time quantitative imaging approach.

%

\ifCLASSOPTIONcaptionsoff
  \newpage
\fi

\bibliographystyle{IEEEtran}
\bibliography{bare_jrnl.bib}

\begin{thebibliography}{10}
\providecommand{\url}[1]{#1}
\csname url@samestyle\endcsname
\providecommand{\newblock}{\relax}
\providecommand{\bibinfo}[2]{#2}
\providecommand{\BIBentrySTDinterwordspacing}{\spaceskip=0pt\relax}
\providecommand{\BIBentryALTinterwordstretchfactor}{4}
\providecommand{\BIBentryALTinterwordspacing}{\spaceskip=\fontdimen2\font plus
\BIBentryALTinterwordstretchfactor\fontdimen3\font minus
  \fontdimen4\font\relax}
\providecommand{\BIBforeignlanguage}[2]{{%
\expandafter\ifx\csname l@#1\endcsname\relax
\typeout{** WARNING: IEEEtran.bst: No hyphenation pattern has been}%
\typeout{** loaded for the language `#1'. Using the pattern for}%
\typeout{** the default language instead.}%
\else
\language=\csname l@#1\endcsname
\fi
#2}}
\providecommand{\BIBdecl}{\relax}
\BIBdecl

\bibitem{pastorino_book}
M.~Pastorino, \emph{{Microwave Imaging}}.\hskip 1em plus 0.5em minus
  0.4em\relax John Wiley \& Sons, 2010.

\bibitem{sandhu_pasd}
A.~I. Sandhu, A.~Desmal, and H.~Bagci, ``{An accelerated nonlinear contrast
  source inversion scheme for sparse electromagnetic imaging},'' \emph{IEEE
  Access}, vol.~9, pp. 54\,811--54\,819, 2021.

\bibitem{sandhu_bed}
A.~I. Sandhu \emph{et~al.}, ``{Bayesian Experimental Design for Efficient
  Sensor Placement in Two-Dimensional Electromagnetic Imaging},'' \emph{IEEE
  Access}, vol.~11, 2023.

\bibitem{D4}
L.~Li \emph{et~al.}, ``{Deepnis: Deep neural network for nonlinear
  electromagnetic inverse scattering},'' \emph{IEEE Trans. Antennas Propag.},
  vol.~67, no.~3, pp. 1819--1825, 2018.

\bibitem{D5}
Z.~Wei and X.~Chen, ``{Deep-learning schemes for full-wave nonlinear inverse
  scattering problems},'' \emph{IEEE Trans. Geosci. Remote Sens.}, vol.~57,
  no.~4, pp. 1849--1860, 2018.

\bibitem{D7}
J.~Li \emph{et~al.}, ``{3-D voxel-based reconstruction of multiple objects
  buried in layered media by vbim hybridized with unsupervised machine
  learning},'' \emph{IEEE Trans. Geosci. Remote Sens.}, 2021.

\bibitem{A28}
S.~Qi \emph{et~al.}, ``{Two-dimensional electromagnetic solver based on deep
  learning technique},'' \emph{IEEE Journal on Multiscale and Multiphysics
  Computational Techniques}, vol.~5, pp. 83--88, 2020.

\bibitem{A29}
Z.~Ma \emph{et~al.}, ``{Learning-based fast electromagnetic scattering solver
  through generative adversarial network},'' \emph{IEEE Trans. Antennas
  Propag.}, 2020.

\bibitem{backpropag_cnn}
Z.~Wei and X.~Chen, ``{Deep-Learning Schemes for Full-Wave Nonlinear Inverse
  Scattering Problems},'' \emph{IEEE Transactions on Geoscience and Remote
  Sensing}, vol.~57, no.~4, pp. 1849--1860, April 2019.

\bibitem{sandhu_cosamp}
A.~I. Sandhu, S.~A. Shaukat, A.~Desmal, and H.~Bagci, ``{ANN-assisted CoSaMP
  algorithm for linear electromagnetic imaging of spatially sparse domains},''
  \emph{IEEE Trans. Antennas Propag.}, vol.~69, no.~9, pp. 6093--6098, 2021.

\bibitem{D}
A.~Desmal, ``{A Trained Iterative Shrinkage Approach Based on Born Iterative
  Method for Electromagnetic Imaging},'' \emph{IEEE Transactions on Microwave
  Theory and Techniques}, vol.~70, no.~11, pp. 4991--4999, Nov. 2022.

\bibitem{A}
R.~Guo \emph{et~al.}, ``{Physics Embedded Deep Neural Network for Solving
  Volume Integral Equation: 2-D Case},'' \emph{IEEE Trans. Antennas Propag.},
  vol.~70, no.~8, pp. 6135--6147, Aug. 2022.

\bibitem{GANpaper}
Z.~Ma \emph{et~al.}, ``{Learning-Based Fast Electromagnetic Scattering Solver
  Through Generative Adversarial Network},'' \emph{IEEE Trans. Antennas
  Propag.}, vol.~69, no.~4, pp. 2194--2208, April 2021.

\bibitem{GAN36}
P.~Isola, J.~Y. Zhu, T.~Zhou, and A.~A. Efros, ``{Image-to-image translation
  with conditional adversarial nets},'' 2017.

\bibitem{GAN37}
I.~Goodfellow, J.~P.-A. adie, M.~Mirza, B.~Xu, D.~Warde-Farley, S.~Ozair,
  A.~Courville, and Y.~Bengio, ``{Generative adversarial nets},'' pp.
  2672--2680, 2014.

\bibitem{D1}
L.-Y. Xiao, J.~Li, F.~Han, W.~Shao, and Q.~H. Liu, ``{Dual-module NMM-IEM
  machining learning for fast electromagnetic inversion of inhomogeneous
  scatterers with high contrasts and large electrical dimensions},'' \emph{IEEE
  Trans. Antennas Propag.}, 2020.

\bibitem{D2}
H.~M. Yao, E.~Wei, and L.~Jiang, ``{Two-step enhanced deep learning approach
  for electromagnetic inverse scattering problems},'' \emph{IEEE Antennas
  Wireless Propag. Lett.}, vol.~18, no.~11, pp. 2254--2258, 2019.

\bibitem{D3}
R.~Mydur and K.~A. Michalski, ``{A neural-network approach to the
  electromagnetic imaging of elliptic conducting cylinders},'' \emph{Microw.
  and Opt. Technol. Lett.}, vol.~28, no.~5, pp. 303--306, 2001.

\bibitem{W18}
W.~Ma \emph{et~al.}, ``{Deep learning for the design of photonic structures},''
  \emph{Nature Photonics}, vol.~15, no.~2, pp. 77--90, 2021.

\bibitem{W19}
C.~Gurbuz \emph{et~al.}, ``{Generative adversarial networks for the design of
  acoustic metamaterials},'' \emph{The Journal of the Acoustical Society of
  America}, vol. 149, no.~2, pp. 1162--1174, 2021.

\bibitem{W20}
W.~Ma \emph{et~al.}, ``{Probabilistic representation and inverse design of
  metamaterials based on a deep generative model with semi‐supervised
  learning strategy},'' \emph{Advanced Materials}, vol.~31, no.~35, 2019.

\bibitem{ahmed2021deterministic}
W.~W. Ahmed, M.~Farhat, X.~Zhang, and Y.~Wu, ``Deterministic and probabilistic
  deep learning models for inverse design of broadband acoustic cloak,''
  \emph{Physical Review Research}, vol.~3, no.~1, p. 013142, 2021.

\bibitem{ahmed2023machine}
W.~W. Ahmed, M.~Farhat, K.~Staliunas, X.~Zhang, and Y.~Wu, ``Machine learning
  for knowledge acquisition and accelerated inverse-design for non-hermitian
  systems,'' \emph{Communications Physics}, vol.~6, no.~1, p.~2, 2023.

\bibitem{W21}
P.~Meng \emph{et~al.}, ``{Solving a kind of inverse scattering problem of
  acoustic waves based on linear sampling method and neural network},''
  \emph{Alexandria Engineering Journal}, vol.~59, no.~3, pp. 1451--1462, 2020.

\bibitem{W24}
Y.~T. Luo \emph{et~al.}, ``{Probability-density-based deep learning paradigm
  for the fuzzy design of functional metastructures},'' \emph{Research}, 2020.

\bibitem{W25}
W.~W. Ahmed \emph{et~al.}, ``{Deterministic and probabilistic deep learning
  models for inverse design of broadband acoustic cloak},'' \emph{Physical
  Review Research}, vol.~3, no.~1, 2021.

\bibitem{W26}
T.~Tran \emph{et~al.}, ``{Broadband Acoustic Metamaterial Design via Machine
  Learning},'' \emph{Journal of Theoretical and Computational Acoustics},
  vol.~30, no.~03, 2022.

\bibitem{W27}
Z.~Liu \emph{et~al.}, ``{Generative model for the inverse design of
  metasurfaces},'' \emph{Nano letters}, vol.~18, no.~10, pp. 6570--6576, 2018.

\bibitem{W29}
P.~Lai, F.~Amirkulova, and P.~Gerstoft, ``{Conditional Wasserstein generative
  adversarial networks applied to acoustic metamaterial design},'' \emph{The
  Journal of the Acoustical Society of America}, vol. 150, no.~6, pp.
  4362--4374, 2021.

\bibitem{W}
W.~W. Ahmed \emph{et~al.}, ``{A Generative deep learning approach for shape
  recognition of arbitrary objects from phaseless acoustic scattering data},''
  \emph{Advanced Intelligent Systems}, vol.~5, no.~5, 2023.

\bibitem{richmond1965}
J.~Richmond, ``{Scattering by a dielectric cylinder of arbitrary cross section
  shape},'' \emph{IEEE Trans. Antennas Propag.}, vol.~13, no.~3, pp. 334--341,
  1965.

\bibitem{W37}
D.~P. Kingma and M.~Welling, ``{Auto-encoding variational bayes},'' \emph{arXiv
  preprint arXiv:1312.6114}, 2013.

\bibitem{W36}
I.~Goodfellow \emph{et~al.}, ``{Generative adversarial nets},'' vol.~27, 2014.

\end{thebibliography}
\end{document}